\documentclass[amsmath,amssymb,floatfix,lengthcheck,preprintnumbers,prl,showpacs,superscriptaddress,nofootinbib,twocolumn,aps]{revtex4-1}
\usepackage{graphicx}
\usepackage{slashed}
\usepackage{xcolor}
\usepackage{mathtools}

\newcommand{\gsim}{\mathrel{\raisebox{-.6ex}{$\stackrel{\textstyle>}{\sim}$}}}

\long\def\/*#1*/{}
\definecolor{red}{rgb}{1.0, 0, 0}

\begin{document}

\title{Direct Detection of Stealth Dark Matter through Electromagnetic Polarizability}

\author{T.~Appelquist}
\affiliation{Department of Physics, Sloane Laboratory, Yale University, New Haven, Connecticut 06520, USA}
\author{E.~Berkowitz}
\affiliation{Lawrence Livermore National Laboratory, Livermore, California 94550, USA}
\author{R.~C.~Brower}
\affiliation{Department of Physics, Boston University,
	Boston, Massachusetts 02215, USA}
\author{M.~I.~Buchoff}
\affiliation{Institute for Nuclear Theory, Box 351550, Seattle, WA 98195-1550, USA}
\author{G.~T.~Fleming}
\affiliation{Department of Physics, Sloane Laboratory, Yale University, New Haven, Connecticut 06520, USA}
\author{X.-Y.~Jin}
\affiliation{Argonne Leadership Computing Facility, Argonne, Illinois 60439, USA}
\author{J.~Kiskis}
\affiliation{Department of Physics, University of California, Davis, California 95616, USA}
\author{G.~D.~Kribs}
\affiliation{Department of Physics, University of Oregon, Eugene, OR, 97403 USA}
\author{E.~T.~Neil}
\affiliation{Department of Physics, University of Colorado, Boulder, CO 80309, USA}
\affiliation{RIKEN-BNL Research Center, Brookhaven National Laboratory, Upton, NY 11973, USA}
\author{J.~C.~Osborn}
\affiliation{Argonne Leadership Computing Facility, Argonne, Illinois 60439, USA}
\author{C.~Rebbi}
\affiliation{Department of Physics, Boston University, Boston, Massachusetts 02215, USA}
\author{E.~Rinaldi}
\affiliation{Lawrence Livermore National Laboratory, Livermore, California 94550, USA}
\author{D.~Schaich}
\affiliation{Department of Physics, Syracuse University, Syracuse, NY 13244, USA}
\author{C.~Schroeder}
\affiliation{Lawrence Livermore National Laboratory, Livermore, California 94550, USA}
\author{S.~Syritsyn}
\affiliation{RIKEN-BNL Research Center, Brookhaven National Laboratory, Upton, NY 11973, USA}
\author{P.~Vranas}
\affiliation{Lawrence Livermore National Laboratory, Livermore, California 94550, USA}
\author{E.~Weinberg}
\affiliation{Department of Physics, Boston University, Boston, Massachusetts 02215, USA}
\author{O.~Witzel}
\altaffiliation[Present address: ]{Higgs Centre for Theoretical Physics,
School
of Physics \& Astronomy, The University of Edinburgh, EH9 3FD, UK}
\affiliation{Center for Computational Science, Boston University, Boston, MA,
USA}
\collaboration{Lattice Strong Dynamics (LSD) Collaboration}

\begin{abstract}

We calculate the spin-independent scattering cross section for direct detection that results from the electromagnetic polarizability of a composite scalar baryon dark matter candidate -- ``Stealth Dark Matter'', based on a dark SU(4) confining gauge theory.  In the nonrelativistic limit, electromagnetic polarizability proceeds through a dimension-7 interaction leading to a very small scattering cross section for dark matter with weak-scale masses. This represents a lower bound on the scattering cross section for composite dark matter theories with electromagnetically charged constituents.  We carry out lattice calculations of the polarizability for the lightest baryons in SU(3) and SU(4) gauge theories using the background field method on quenched configurations.  We find the polarizabilities of SU(3) and SU(4) to be comparable (within about 50\%) normalized to the baryon mass, which is suggestive for extensions to larger SU(N) groups. The resulting scattering cross sections with a xenon target are shown to be potentially detectable in the dark matter mass range of about 200-700~GeV, where the lower bound is from the existing LUX constraint while the upper bound is the coherent neutrino background. Significant uncertainties in the cross section remain due to the more complicated interaction of the polarizablity operator with nuclear structure, however the steep dependence on the dark matter mass,  $1/m_B^6$, suggests the observable dark matter mass range is not appreciably modified. We briefly highlight collider searches for the mesons in the theory as well as the indirect astrophysical effects that may also provide excellent probes of stealth dark matter.

\end{abstract}

\pacs{11.15.Ha, 12.60.-i, 95.35.+d}
\preprint{INT-PUB-15-005, LLNL-JRNL-667121}

\maketitle

\paragraph{Introduction --} Despite remarkable advances in direct detection experiments~\cite{Akerib:2013tjd,Agnese:2014aze,Cushman:2013zza,Xiao:2014xyn}, a conclusive signal of nuclear interactions with dark matter (DM) remains elusive. These experiments, which are sensitive to nucleus-DM cross sections at or below $10^{-45}\ \text{cm}^2$ per nucleon, have already excluded large classes of interactions and models, and are now actively probing Higgs boson exchange~\cite{Cushman:2013zza,Xiao:2014xyn}.  

Composite DM, which arises as a neutral bound state of a strongly-coupled gauge force (for early work, see \cite{Nussinov:1985xr,Chivukula:1989qb,Barr:1990ca,Barr:1991qn,Kaplan:1991ah}), has sparked multiple recent lattice calculations  \cite{Lewis:2011zb,Hietanen:2012sz, Appelquist:2013ms, Hietanen:2013fya, Appelquist:2014jch,Detmold:2014qqa,Detmold:2014kba, Fodor:2015eea}.  If its constituents are electromagnetically charged, the DM will interact with standard model (SM) nuclei via photon exchange, with the cross section suppressed by a momentum-dependent electromagnetic form factor.  Expanding at small momentum transfer, one can obtain a series of effective operators describing the interaction: the dimension-5 magnetic moment, dimension-6 charge radius, and dimension-7 polarizability are the leading operators \cite{Chivukula:1992pn,Bagnasco:1993st,Pospelov:2000bq}.  Symmetry considerations can give models in which the first two operators are identically zero \cite{Kribs:2009fy,Buckley:2012ky}. Scattering due to electromagnetic polarizability remains, giving a lower bound on the direct detection cross section for a composite DM particle with charged constituents.  

We will focus here on a particular composite DM model, ``stealth dark matter'' \cite{Appelquist:2015yfa}, in which the DM is a scalar baryon composed of dark fermions that transform under an SU$(N_D)$ theory with $N_D$ even.  In this model, electroweak symmetry breaking proceeds through the SM Higgs mechanism. The dominant modes of interaction with the SM are the polarizability operator and direct Higgs boson exchange.  The latter was studied in some detail in \cite{Appelquist:2014jch,Appelquist:2015yfa}, placing bounds on the allowed dark matter coupling to the Higgs boson.  In this work we study the polarizability, which unlike the Higgs interaction has no adjustable parameters, but rather is completely determined by the strong dynamics once the gauge group and matter content are specified.

\paragraph{SU(4) baryons and polarizabilities --} A full construction of the stealth DM model is given in \cite{Appelquist:2015yfa}; here we briefly summarize the relevant details.  The dark sector consists of an unbroken SU(4) gauge theory, which contains bosonic baryonic bound states made up of four constituent fermions.  The DM candidate itself is a scalar made up of two pairs of fermions which are degenerate in mass and carry equal but opposite electric charges of $\pm 1/2$. Hence, there is no magnetic moment or charge radius, leaving just the electromagnetic polarizability as the dominant interaction with photons.

Previous estimates of the polarizability of a composite scalar have led to direct-detection cross sections on the order of $10^{-48}\ \text{cm}^2$~\cite{Pospelov:2000bq}, approaching the interaction strength at which background neutrinos are expected to contaminate the DM recoil signal. However, the estimates were based on semi-classical calculations of a strongly-coupled interaction, and so have uncontrolled uncertainties.

Additionally, due to how internal electric charges are correlated, the polarizability of bosonic 4-fermion baryons may differ appreciably from QCD-based estimates.  In one limit where the internal constituents are uncorrelated, the polarizabilities are expected to be comparable. However, if alternate flavors tend to form pairs based on their Pauli statistics, the 4-fermion baryon polarizability would be derivative-suppressed compared to the 3-fermion baryon (i.e.~two dipoles vs.\ one dipole and one charge). 
In order to quantify this effect, we perform lattice calculations for both the SU(3) and SU(4) baryon polarizabilities.

\paragraph{Polarizability and Direct Detection --}  The electric polarizability 
of the scalar baryonic composite DM field $B$ with mass $m_B$ 
can be written as an effective operator of the form
\begin{equation}\label{eq:O_F}
\mathcal{O}_{F} = C_F B^* B \, F^{\mu\alpha}F^{\nu}_{\alpha} v_\mu v_\nu
\end{equation}
where $F_{\mu \nu}$ is the electromagnetic field strength tensor, 
$v_\mu = (1, 0, 0, 0)$ in the static limit, 
and $C_F$ is the polarizability with 
mass dimension $-3$ in the nonrelativistic limit. 
Only the electric polarizability is considered since the 
magnetic polarizability is expected to be suppressed \cite{Luke:1992tm}.
This is a two-photon vertex, so that the scattering off of nuclei will involve a virtual photon loop.  Because this operator is induced at a high scale (roughly the dark confinement scale $\Lambda_D \sim m_B$), it is expected to generate other interactions with SM particles when the appropriate effective field theory matching and running down to the nuclear scale are carried out \cite{Fitzpatrick:2012ib, Fitzpatrick:2012ix, Hill:2014yka, Hill:2014yxa}; in fact, an explicit treatment for the polarizability operator is given in \cite{Frandsen:2012db}.  Although the effects of the additional induced operators are not negligible in general, we find that they are small compared to the uncertainties (particularly from nuclear physics) 
and so we will omit them. 

From the interaction shown above, the coherent DM-nucleus scattering cross section (per nucleon) is given by 
\begin{equation}\label{eq:cross_section}
\sigma_{\rm nucleon}(Z, A) = \frac{\mu_{n B}^2}{\pi A^2}\left\langle\left| C_F f_F^A \right|^2\right\rangle,
\end{equation}
where $m_n$ is the nucleon mass, $\mu_{n B}= m_n m_B/(m_n + m_B)$ is the reduced mass, ($Z$, $A$) are the atomic and mass numbers of the target nucleus, and the angular brackets represent the momentum-averaged form factors for heavy DM candidates in a given experiment \cite{Frandsen:2012db}.

The primary source of systematic uncertainty is on the nuclear physics side of the calculation -- evaluating the non-perturbative nuclear matrix element, $f_F^A = \langle A| F^{\mu\nu}F_{\mu\nu} | A \rangle$.   
Various attempts to perturbatively estimate this matrix element have been performed with varying levels of complexity~\cite{Weiner:2012cb, Frandsen:2012db, Ovanesyan:2014fha}.  But, the matrix element also has nontrivial excited-state structures that likely require a fully non-perturbative treatment.  This matrix element is similar to those needed for double-beta decay experiments, estimates for which have substantial variation~\cite{Kotila:2012zza,Mustonen:2013zu}. Until a more accurate extraction of this matrix element is performed, we will use dimensional analysis arising from non-relativistic loop momenta counting,
\begin{equation}\label{eq:nuclear-naive-scaling}
f_F^A \sim 3 Z^2 \alpha \frac{M_F^A}{R},
\end{equation}
where $R = 1.2 A^{1/3}\ \text{fm}$, as used in the double beta decay context, $\alpha$ is the fine-structure constant, and $M_F^A$ is a dimensionless parameter.  With the factor of $3$ in Eq.~(\ref{eq:nuclear-naive-scaling}),   our expression approximately matches \cite{Weiner:2012cb,Frandsen:2012db} for heavy nuclei when $M_F^A \simeq 1$.  To allow for an order of magnitude uncertainty in the nuclear matrix element, we take $1/3 < M_F^A < 3$, although a detailed nuclear structure extraction would be needed for a more precise estimate.

\paragraph{Background field method --}

In order to extract the electric polarizability from the lattice, the background field method is employed, as described in Ref.~\cite{Detmold:2009dx,Detmold:2010ts}. The essence of this method is to measure baryon two-point correlation functions in the presence of a uniform electric field $\mathcal{E}$.  Working in Euclidean space, the background field induces a quadratic Stark shift in the mass of the SU$(4)$ ground-state baryon,
\begin{equation}\label{eq:Stark4c}
E_{B,4c} = m_B + 2C_F |\mathcal{E}|^2 + \mathcal{O}\left(\mathcal{E}^4\right)\ ,
\end{equation}
where $C_F$ is the desired polarizability\footnote{The electric polarizability of the neutron $\alpha_E$ is more commonly defined in terms of the induced dipole moment $\vec{d} = 4\pi \alpha_E \vec{\mathcal{E}}$, giving a quadratic Stark shift of $\Delta E_n = \frac{1}{2} \vec{d} \cdot \vec{\mathcal{E}} = 2\pi \alpha_E |\mathcal{E}|^2$.  In our notation $\alpha_E = C_F / \pi$.}, as defined in Eq.~(\ref{eq:O_F}).  

Due to the scalar nature of the SU$(4)$ baryon ground state, this relation is equivalent to what one would expect for mesons. For comparison we also study  the fermionic SU(3) baryon, whose energy shift contains an additional contribution from the non-zero magnetic moment $\mu_B$~\cite{Detmold:2010ts},
\begin{equation}\label{eq:Stark3c}
{E}_{B,3c} = {m}_B + \left(2{C}_F- \frac{{\mu_B}^2}{8{m}_B^3}\right){|\mathcal{E}|}^2 + \mathcal{O}\left({\mathcal{E}}^4\right).
\end{equation}
For the SU(3) theory, we must therefore determine $\mu_B$ as well in order to extract $C_F$ from the background field dependence.

The background field method is implemented following Refs.~\cite{Detmold:2009dx,Detmold:2010ts} where the uniform background field is included by multiplying the unitary gauge links by two phase terms, chosen so that the field is oriented in the $\hat{z}$ direction.  Quantization of the uniform background field on a torus restricts the available field strengths to values
\begin{equation}\label{eq:quant_E}
\mathcal{E}/a^2 =\frac{2\pi e n}{|q_{\text{low}}| N_t N_s},
\end{equation}
where $a$ is the lattice spacing, $e$ is the electromagnetic coupling, $N_s$ and $N_t$ are the number of spatial and temporal lattice sites respectively, and $q_{\text{low}}$ is the lowest common denominator of the charges (for SU(3), $q_{\text{low}}=1/3$; for SU(4), $q_{\text{low}}=1/2$).  

For convenience we define a rescaled, dimensionless background field by $\tilde{\mathcal{E}} = (ea^2)^{-1} \mathcal{E}$. We will analyze our lattice results using Eqs.~(\ref{eq:Stark4c})-(\ref{eq:Stark3c}) with all quantities replaced with their rescaled, dimensionless counterparts, all of which will be denoted with a tilde:
\begin{align}
m_B &= \tilde{m}_B / a, \\
C_F &= 4\pi \alpha a^3 \tilde{C}_F, \label{eq:cf_phys} \\
\mu_B &= 4\pi \alpha \tilde{\mu}_B.
\end{align}
%

\paragraph{Lattice details and fitting --} The lattice calculations are done using the \textit{Chroma} software package~\cite{Edwards:2004sx}.  We use the plaquette gauge action with unimproved Wilson fermions.  The gauge configurations are quenched $N_s^3 \times N_t = 32^3 \times 64$ lattices (20000 heat-bath updates, 200 configurations separated evenly). For SU(4) we choose $\beta=11.028$ and for SU(3) $\beta=6.0175$ following \cite{Bali:2008an}.  Fermionic propagators are calculated for two different masses at each $N_D$ value ($\kappa=0.1554,0.15625$ for $N_D=4$ and $\kappa=0.1537,0.1547$ for $N_D=3$), chosen such that the ratio of the pseudoscalar to vector meson masses  $m_{PS}/m_V = 0.77$ and $0.70$ are matched~\cite{Bali:2008an, Appelquist:2014jch}.

Background field measurements are performed at six field values [$n=0,...,5$, see Eq.~(\ref{eq:quant_E})] for both $N_D=4$ and $N_D=3$, with correlation functions measured using 40 evenly separated sources in $(x,y)$ along the $t=z=0$ plane. Each zero and non-zero field value has 8000 measurements.  All two-point correlation functions are fit over the range $t \in [4,28]$ using fully correlated, multi-exponential fits including three excited states.

For $N_D = 4$, the two-point baryonic correlation function in background field $\tilde{\mathcal{E}}$ takes the form 
\begin{equation}
C_B(t, \tilde{\mathcal{E}}) \sim Z(\tilde{\mathcal{E}}) \exp \left[-t \tilde{E}_B(\tilde{\mathcal{E}})\right]
\end{equation}
at large $t$.  Results for $\tilde{E}_B(\tilde{\mathcal{E}})$ are then fit to Eq.~(\ref{eq:Stark4c}).  We include higher-order contributions from the background field following \cite{Detmold:2010ts}, 
\begin{equation}\label{eq:quadCF}
\tilde{C}_F(\tilde{\mathcal{E}}) = \tilde{C}_F + \tilde{C}_F' |\tilde{\mathcal{E}}|^2.
\end{equation}

The analysis for $N_D = 3$ is complicated by the contribution of the magnetic moment $\tilde{\mu}_B$ to the baryon self-energy.  Following \cite{Detmold:2010ts}, we make use of the boost projections
\begin{equation}
\mathcal{P}_\pm = \frac{1}{2}\left(1\pm i\gamma_3\gamma_4\right),
\end{equation}
and the boosted correlators
\begin{eqnarray}
C_B^{\pm}(t) &=& \langle 
\bar{B}(\textbf{x},t) \mathcal{P}_\pm B(\textbf{0},0) \rangle_\mathcal{E} \nonumber \\
&=& Z_{\pm} (\tilde{\mathcal{E}}) \exp\left[-t \tilde{E}_B(\tilde{\mathcal{E}})\right].
\end{eqnarray}
The boost-projected amplitudes $Z_{\pm}$ contain equal and opposite contributions from the magnetic moment, which we isolate by combining them in the ratio
\begin{equation}\label{eq:Zratio}
Z_r \equiv \frac{Z_+(\tilde{\mathcal{E}}) - Z_-(\tilde{\mathcal{E}})}{Z_+(\tilde{\mathcal{E}}) + Z_-(\tilde{\mathcal{E}})} = \frac{\tilde{\mathcal{E}} \tilde{\mu}_B(\tilde{\mathcal{E}})}{2\tilde{m}_B^2}.
\end{equation}
A simultaneous fit of $\tilde{E}_B$ to Eq.~(\ref{eq:Stark3c}) and the amplitude ratio in Eq.~(\ref{eq:Zratio}) allows us to determine both $\tilde{C}_F$ and $\tilde{\mu}_B$.  To extract the polarizability $\tilde{C}_F$ we use a fully correlated quadratic fit following Eqs.~(\ref{eq:Stark4c})-(\ref{eq:Stark3c}).  Once again we incorporate  quadratic terms to both $\tilde{C}_F$ [as in Eq.~(\ref{eq:quadCF})] and $\tilde{\mu}_B$,
\begin{equation}
\tilde{\mu}_B(\tilde{\mathcal{E}}) = \tilde{\mu}_B + \tilde{\mu}_B' |\tilde{\mathcal{E}}|^2.
\end{equation}
%

\begin{figure}[t]
\begin{center}
\includegraphics[width=0.48\textwidth]{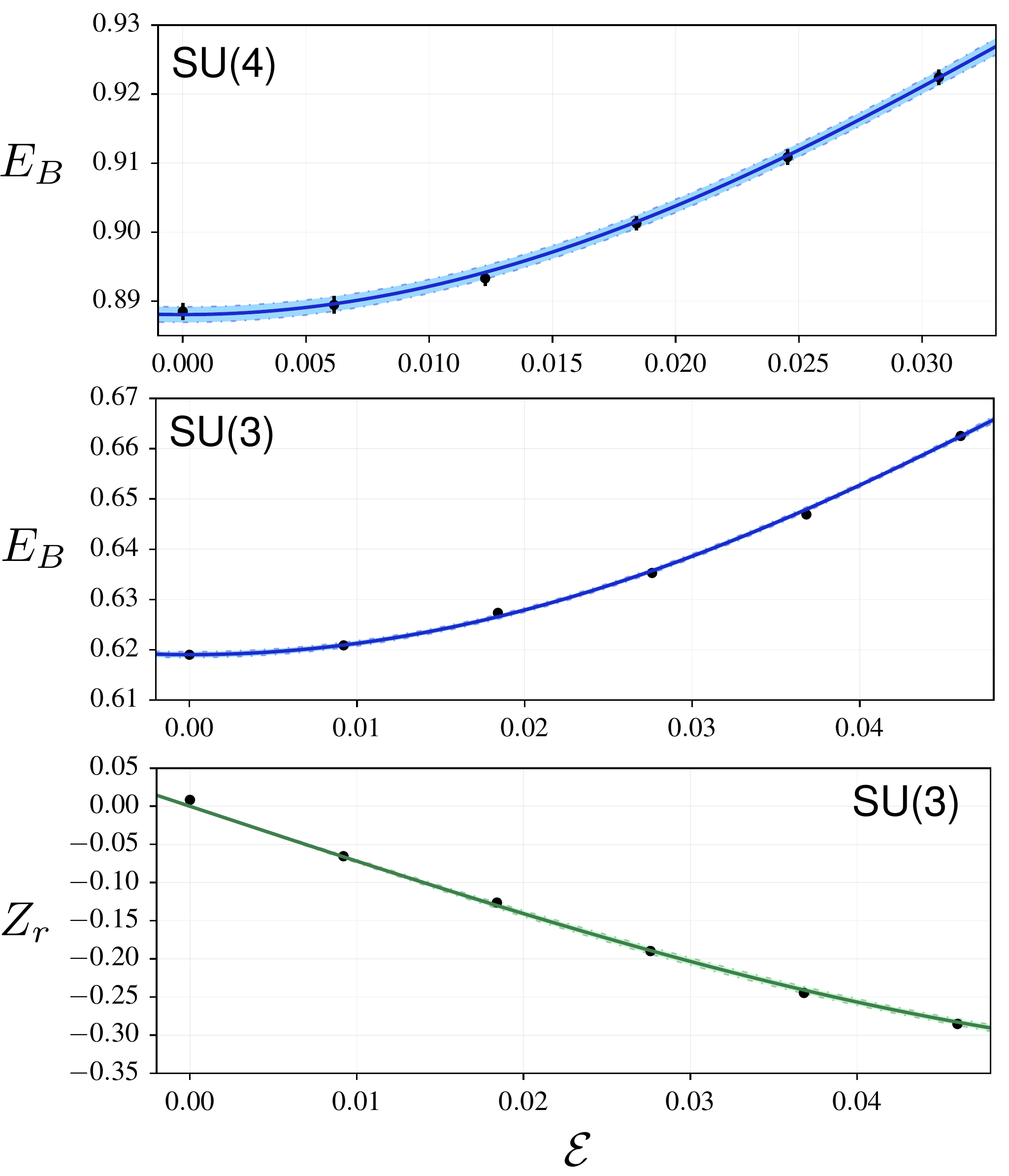}
\end{center}
\caption{The ground state energy (in lattice units) vs. applied electric field $\mathcal{E}$ for SU(4) baryons (top) and SU(3) baryons (middle), and ratio of projected SU(3) correlators vs. $\mathcal{E}$ (bottom). Their relations to the magnetic moment and polarizabilities are presented in Eqs.~(\ref{eq:Stark4c}), (\ref{eq:Stark3c}) and (\ref{eq:Zratio}).  Results shown are for the ensembles with $m_{PS} / m_V = 0.70$.}
\label{fig:Pol_Fit_4c}
\end{figure}

The polarizability results for both SU(4) and SU(3) are presented in Table~\ref{tab:results}, and results for the energies and the ratio $Z_r$ vs.\ background field for the $m_{PS} / m_V = 0.70$ ensembles are plotted in Fig.~\ref{fig:Pol_Fit_4c}.

Constructing the dimensionless product $\alpha \tilde{C}_F \tilde{m}_B^3$ (as needed for the DM cross section), we find that the SU(4) polarizabilities are  larger than SU(3) by about $50\%$. Thus, we find the SU(3) and SU(4) polarizabilities to be comparable when normalized to the baryon mass. Of course, the baryon mass itself scales proportional to $N_D$; if we were to set the scale using a quantity such as the string tension which does not scale with $N_D$, then the SU(3) polarizability would be larger.

\begin{table}[t]
 \centering
 {\scriptsize
 \begin{tabular}{|c|c|c|c|c|c|c|c|}
 \hline
  $N_D$& $m_{PS}/m_V$ & $\tilde{m}_B$ &$\alpha \tilde{C}_F$ & $\alpha^2 \tilde{C}'_F$ &$\tilde{\mu}_B$&$\tilde{\mu}'_B$ & $\chi^2/\text{dof}$\\
 \hline
 4 & 0.77 & 0.98204(93) & 0.1420(56) & -0.089(29) & --- & --- & 0.7/3 \\
 \hline
 & 0.70 & 0.88805(113)& 0.1514(106) & -0.142(68) &  --- & --- & 4.8/3\\
 \hline
 3 & 0.77 & 0.69812(51) &0.2829(127) & -0.177(45) & -6.87(26) & 714(103) & 3.0/7\\
 \hline
 & 0.70 & 0.61904(59) &0.2829(81) & -0.165(24) & -5.55(18) & 396(78) & 13.4/7 \\
 \hline
 \end{tabular} }
 \caption{Results for the polarizabilities and magnetic moments of the baryonic composites of a strongly-coupled SU($N_D$) theory, in lattice units.}
 \label{tab:results}
\end{table}

The effect of the quenched approximation, in which 
dynamical fermion loops are omitted from the lattice calculation, 
is not straightforward to estimate.  However, the effects 
of such loops are expected to be suppressed with large $N_D$ 
and heavy fermion mass; we note that even for QCD with its 
much lighter fermions, the effects of quenching are generally 
at most of order $10\%$ \cite{Colangelo:2010et}.

Our calculations are  performed at a single lattice spacing 
and volume, both of which can lead to additional systematic effects.  
We expect all of these corrections to be small relative to 
the order of magnitude uncertainty taken for the 
nuclear matrix element $M_F^A$.  
As a cross-check, we note that the neutron polarizability 
from the PDG \cite{Agashe:2014kda} gives 
$C_F m_n^3 \simeq 0.36$ at the QCD physical ratio $m_{PS} / m_V = 0.18$, 
while our SU(3) lattice simulations give 
$C_F m_B^3 \simeq 0.84$ at $m_{PS} / m_V = 0.70$.  
These results are broadly consistent with the expected scaling 
of the polarizability and baryon mass with $m_{PS}$.

\paragraph{Direct detection cross sections --} To relate the dimensionless lattice results to the dimensionful DM mass, $m_B$, that we vary continuously in order to scan the parameter space of the theory, it is most convenient to give units to the lattice spacing $a = \tilde{m}_B/m_B$. Along with Eq.~(\ref{eq:cf_phys}), this leads to the physical value of the polarizability
\begin{equation}
C_F = 4\pi \alpha\left(\frac{ \tilde{m}_B}{m_B}\right)^3\tilde{C}_F\ .
\end{equation}

\begin{figure}[t]
\begin{center}
\includegraphics[width=0.45\textwidth]{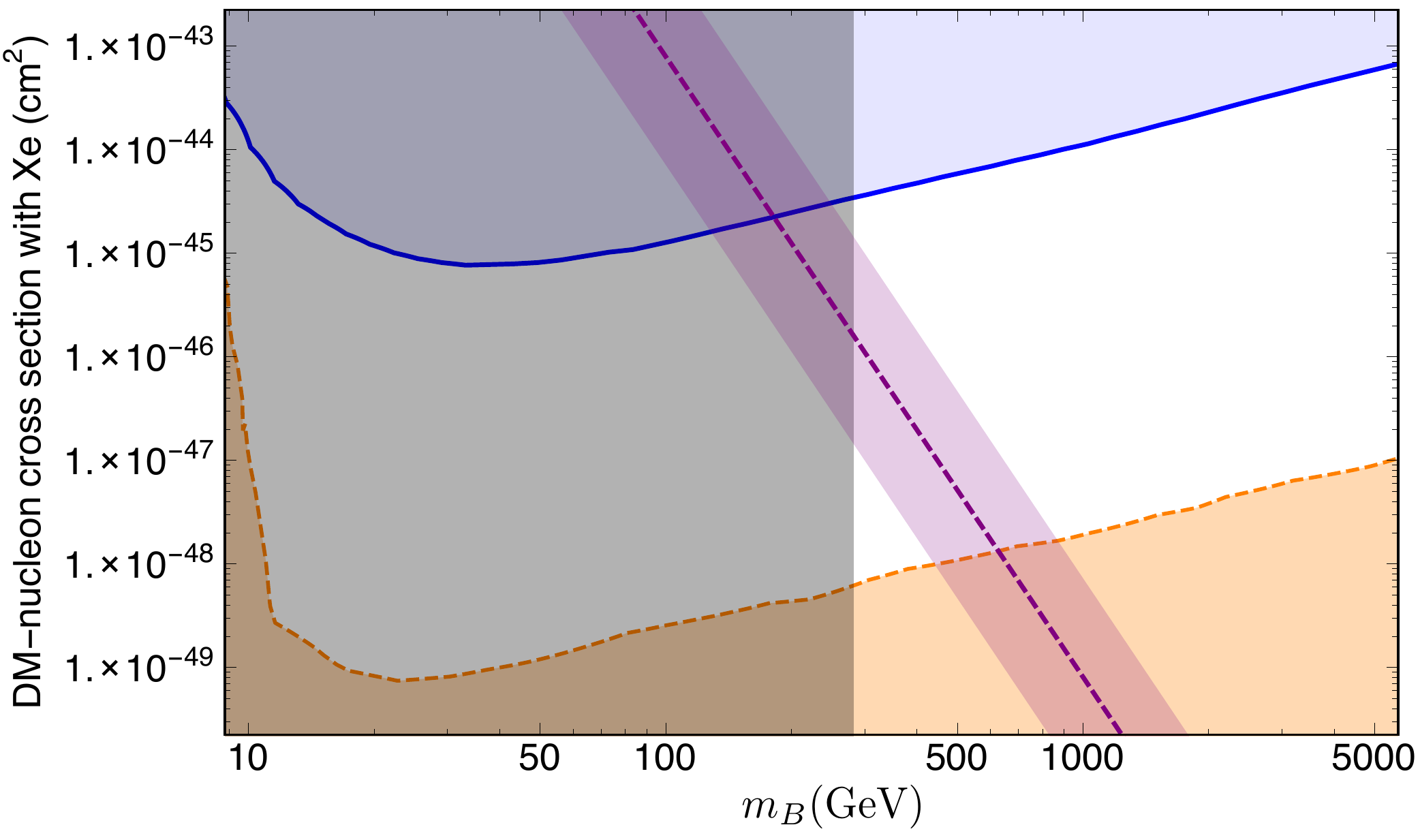}
\end{center}
\caption{The DM spin-independent scattering cross section per nucleon 
evaluated for xenon is shown as the purple band obtained from the 
SU(4) polarizability, where the width of the band corresponds to 
$1/3 < M_F^A < 3$ from low to high. 
The blue curve and the light blue region above it is excluded
by the LUX constraints~\cite{Akerib:2013tjd}.
The vertical, darker shaded region is excluded by the 
LEP II bound on charged mesons \cite{Appelquist:2015yfa}. 
The orange region represents the limit at which direct detection 
experiments will be unable to discriminate DM events from 
coherent neutrino recoil~\cite{Billard:2013qya}.
We emphasize that this plot is applicable for xenon, 
and would require calculating Eq.~(\ref{eq:final}) to apply 
to other nuclei.}
\label{fig:Pol_Exclusion}
\end{figure}

Putting everything together, the spin-independent cross section 
written as the conventional \emph{per nucleon} rate for a nucleus 
with ($Z$, $A$) from Eq.~(\ref{eq:cross_section}) becomes
\begin{equation}
\sigma_{\rm nucleon}(Z, A) = \frac{Z^4}{A^2} 
\frac{144 \pi \alpha^2 \mu_{n B}^2 (M_F^A)^2}{m_B^6 R^2}
[\alpha \tilde{m}_B^3 \tilde{C}_F]^2 \, , \label{eq:final}
\end{equation}
where we use our lattice results in Table~\ref{tab:results} 
to evalate the factor in square brackets.
We emphasize that, unlike Higgs exchange, the cross section 
\emph{per nucleus} scales as $Z^4$ and not $A^2$, and so the 
cross section \emph{per nucleon} must be calculated for each nucleus 
separately in order to compare with experiment.  
The strongest bound on the spin-independent direct detection 
scattering rate is from LUX~\cite{Akerib:2013tjd}. 
In Fig.~\ref{fig:Pol_Exclusion}, we show
the scattering cross section per nucleon for xenon,
and compare with the LUX bounds. 
We plot only the $N_D=4$ case here, as the $N_D=3$ baryons are 
already excluded up to $\sim 20$~TeV in mass by the LUX bounds 
through their magnetic moments~\cite{Appelquist:2013ms}.

\paragraph{Discussion --} 

Our lattice results have allowed us to 
calculate the spin-independent
scattering cross section of SU(4) stealth DM through polarizability, 
which we compare against the LUX constraints in Fig.~\ref{fig:Pol_Exclusion}. 
We find DM masses less than about $200$~GeV are excluded, 
while the DM mass range $200$-$700$~GeV could be probed by future
experiments before reaching the neutrino background \cite{Billard:2013qya}.
Currently, the strongest lower bound on the DM mass arises
indirectly from the constraints on the lighter 
electrically-charged mesons that can be produced and decay 
promptly in collider experiments.  Using our results \cite{Appelquist:2015yfa}, 
we estimate that DM masses below about $280$~GeV are 
excluded given the LEP II bounds on the pseudoscalar mesons.

It is remarkable that a composite DM particle with a weak-scale mass, 
composed of dark fermions charged under the weak
and electromagnetic interaction, can nevertheless be safe from
both direct detection constraints and the LEP II constraint 
once $m_B \gsim 300$~GeV\@.  This suggests there is a serious 
opportunity for future direct detection experiments to probe the model. 
Given that the scattering 
cross section per nucleon scales as $Z^4/(A^2 R^2)$ 
in Eq.~(\ref{eq:final}), the experiments
with the heaviest nuclei are often more sensitive, i.e., 
xenon is $3.4$ times more sensitive than argon if both
experiments reach the same limit on the (conventional) 
spin-independent scattering per nucleon through Higgs exchange. 

With our lattice calculation of the dark matter polarizability 
in this model, the dominant remaining uncertainty 
stems from the treatment of the non-perturbative 
nuclear matrix element in Eq.~(\ref{eq:cross_section}), 
which is similar to the matrix elements required for 
double beta decay. A significant source of uncertainty is, 
for example, the presence of excited states in Xe-129 and Ge-73 
that have energies of 30 and 15 keV, which will be probed 
by the loop in the cross section calculation 
(typical momenta exchanges are roughly at the MeV scale). 
These resonances could appreciably change the resulting cross section, 
though the steep dependence on the dark matter mass suggests 
only a modest equivalent shift of the DM mass.

The brightest opportunity for stealth dark matter discovery
may fall within the domain of the Large Hadron Collider
(and future colliders). 
Meson phenomenology is very promising, since charged mesons can
be produced through electroweak processes and decay 
completely into SM particles.  In contrast, production of the 
dark matter baryon is rare, since it is considerably heavier 
than the mesons and would have form factor suppression. 
This implies the standard missing energy signals 
that arise from DM production and escape from the detector 
are \emph{rare}.  

Finally, there are many avenues for further investigation of
stealth dark matter, detailed in \cite{Appelquist:2015yfa}.  
One vital issue is to better estimate the abundance.  
In the DM mass regime where stealth DM is detectable at 
direct detection experiments, 
the abundance of stealth dark matter can arise naturally 
from an asymmetric production mechanism \cite{Appelquist:2015yfa}
that was considered long ago 
\cite{Barr:1990ca,Barr:1991qn,Kaplan:1991ah}
and more recently reviewed in \cite{Zurek:2013wia}.  
If there is indeed an 
asymmetric abundance of bosonic dark matter, there are 
additional astrophysical consequences
\cite{McDermott:2011jp,Bramante:2013hn,Bertoni:2013bsa} 
that warrant further investigation to constrain or probe stealth DM.

\paragraph{Acknowledgements --} 

MIB would like to thank Silas Beane, Paulo Bedaque, Tom Cohen, Jon Engel, Wick Haxton, David B.\ Kaplan, Jerry Miller, Maxim Pospelov, Sanjay Reddy, Martin Savage, Achim Schwenk, and Luca Vecchi for enlightening discussions on the full complexity of the nuclear physics calculation required for an accurate cross section prediction.  We are also indebted to Brian Tiburzi for access to information that allowed for very useful checks of our background field code.  We thank the Lawrence Livermore National Laboratory (LLNL)  Multiprogrammatic and Institutional Computing program for Grand Challenge allocations and time on the LLNL BlueGene/Q (rzuseq and vulcan) supercomputer. We thank LLNL for funding from LDRD~13-ERD-023 ``Illuminating the Dark Universe with PetaFlops Supercomputing''. Computing support for this work was provided in part from the LLNL Institutional Computing Grand Challenge program and from the USQCD Collaboration, which is funded by the Office of Science of the US Department of Energy. MIB thanks the University of Oregon and University of Maryland, College Park for hospitality during the course of this work. 

This work has been supported by the U.~S.~Department of Energy under Grant Nos.\ DE-SC0008669 and DE-SC0009998 (D.S.), DE-SC0010025 (R.C.B., C.R., E.W.), DE-FG02-92ER-40704 (T.A.), DE-SC0011640 (G.D.K.), DE-FG02-00ER41132 (M.I.B.), and Contracts DE-AC52-07NA27344 (LLNL), DE-AC02-06CH11357 (Argonne Leadership Computing Facility), and by the National Science Foundation under Grant Nos. NSF PHY11-00905 (G.F.), OCI-0749300 (O.W.). Brookhaven National Laboratory is supported by the U.~S.~Department of Energy under contract DE-SC0012704.  S.N.S was supported by the Office of Nuclear Physics in the U.~S.~Department of Energy's Office of Science under Contract DE-AC02-05CH11231.

\bibliography{LSD-SU4}

\begin{thebibliography}{44}%
\makeatletter
\providecommand \@ifxundefined [1]{%
 \@ifx{#1\undefined}
}%
\providecommand \@ifnum [1]{%
 \ifnum #1\expandafter \@firstoftwo
 \else \expandafter \@secondoftwo
 \fi
}%
\providecommand \@ifx [1]{%
 \ifx #1\expandafter \@firstoftwo
 \else \expandafter \@secondoftwo
 \fi
}%
\providecommand \natexlab [1]{#1}%
\providecommand \enquote  [1]{``#1''}%
\providecommand \bibnamefont  [1]{#1}%
\providecommand \bibfnamefont [1]{#1}%
\providecommand \citenamefont [1]{#1}%
\providecommand \href@noop [0]{\@secondoftwo}%
\providecommand \href [0]{\begingroup \@sanitize@url \@href}%
\providecommand \@href[1]{\@@startlink{#1}\@@href}%
\providecommand \@@href[1]{\endgroup#1\@@endlink}%
\providecommand \@sanitize@url [0]{\catcode `\\12\catcode `\$12\catcode
  `\&12\catcode `\#12\catcode `\^12\catcode `\_12\catcode `\%12\relax}%
\providecommand \@@startlink[1]{}%
\providecommand \@@endlink[0]{}%
\providecommand \url  [0]{\begingroup\@sanitize@url \@url }%
\providecommand \@url [1]{\endgroup\@href {#1}{\urlprefix }}%
\providecommand \urlprefix  [0]{URL }%
\providecommand \Eprint [0]{\href }%
\providecommand \doibase [0]{http://dx.doi.org/}%
\providecommand \selectlanguage [0]{\@gobble}%
\providecommand \bibinfo  [0]{\@secondoftwo}%
\providecommand \bibfield  [0]{\@secondoftwo}%
\providecommand \translation [1]{[#1]}%
\providecommand \BibitemOpen [0]{}%
\providecommand \bibitemStop [0]{}%
\providecommand \bibitemNoStop [0]{.\EOS\space}%
\providecommand \EOS [0]{\spacefactor3000\relax}%
\providecommand \BibitemShut  [1]{\csname bibitem#1\endcsname}%
\let\auto@bib@innerbib\@empty
\bibitem [{\citenamefont {Akerib}\ \emph {et~al.}(2013)\citenamefont {Akerib}
  \emph {et~al.}}]{Akerib:2013tjd}%
  \BibitemOpen
  \bibfield  {author} {\bibinfo {author} {\bibfnamefont {D.}~\bibnamefont
  {Akerib}} \emph {et~al.} (\bibinfo {collaboration} {LUX Collaboration}),\
  }\href@noop {} {\  (\bibinfo {year} {2013})},\ \Eprint
  {http://arxiv.org/abs/1310.8214} {arXiv:1310.8214 [astro-ph.CO]} \BibitemShut
  {NoStop}%
\bibitem [{\citenamefont {Agnese}\ \emph {et~al.}(2014)\citenamefont {Agnese}
  \emph {et~al.}}]{Agnese:2014aze}%
  \BibitemOpen
  \bibfield  {author} {\bibinfo {author} {\bibfnamefont {R.}~\bibnamefont
  {Agnese}} \emph {et~al.} (\bibinfo {collaboration} {SuperCDMS
  Collaboration}),\ }\href {\doibase 10.1103/PhysRevLett.112.241302} {\bibfield
   {journal} {\bibinfo  {journal} {Phys.Rev.Lett.}\ }\textbf {\bibinfo {volume}
  {112}},\ \bibinfo {pages} {241302} (\bibinfo {year} {2014})},\ \Eprint
  {http://arxiv.org/abs/1402.7137} {arXiv:1402.7137 [hep-ex]} \BibitemShut
  {NoStop}%
\bibitem [{\citenamefont {Cushman}\ \emph {et~al.}(2013)\citenamefont
  {Cushman}, \citenamefont {Galbiati}, \citenamefont {McKinsey}, \citenamefont
  {Robertson}, \citenamefont {Tait} \emph {et~al.}}]{Cushman:2013zza}%
  \BibitemOpen
  \bibfield  {author} {\bibinfo {author} {\bibfnamefont {P.}~\bibnamefont
  {Cushman}}, \bibinfo {author} {\bibfnamefont {C.}~\bibnamefont {Galbiati}},
  \bibinfo {author} {\bibfnamefont {D.}~\bibnamefont {McKinsey}}, \bibinfo
  {author} {\bibfnamefont {H.}~\bibnamefont {Robertson}}, \bibinfo {author}
  {\bibfnamefont {T.}~\bibnamefont {Tait}},  \emph {et~al.},\ }\href@noop {} {\
   (\bibinfo {year} {2013})},\ \Eprint {http://arxiv.org/abs/1310.8327}
  {arXiv:1310.8327 [hep-ex]} \BibitemShut {NoStop}%
\bibitem [{\citenamefont {Xiao}\ \emph {et~al.}(2014)\citenamefont {Xiao} \emph
  {et~al.}}]{Xiao:2014xyn}%
  \BibitemOpen
  \bibfield  {author} {\bibinfo {author} {\bibfnamefont {M.}~\bibnamefont
  {Xiao}} \emph {et~al.} (\bibinfo {collaboration} {PandaX Collaboration}),\
  }\href {\doibase 10.1007/s11433-014-5598-7} {\bibfield  {journal} {\bibinfo
  {journal} {Sci.China Phys.Mech.Astron.}\ }\textbf {\bibinfo {volume} {57}},\
  \bibinfo {pages} {2024} (\bibinfo {year} {2014})},\ \Eprint
  {http://arxiv.org/abs/1408.5114} {arXiv:1408.5114 [hep-ex]} \BibitemShut
  {NoStop}%
\bibitem [{\citenamefont {Nussinov}(1985)}]{Nussinov:1985xr}%
  \BibitemOpen
  \bibfield  {author} {\bibinfo {author} {\bibfnamefont {S.}~\bibnamefont
  {Nussinov}},\ }\href {\doibase 10.1016/0370-2693(85)90689-6} {\bibfield
  {journal} {\bibinfo  {journal} {Phys.Lett.}\ }\textbf {\bibinfo {volume}
  {B165}},\ \bibinfo {pages} {55} (\bibinfo {year} {1985})}\BibitemShut
  {NoStop}%
\bibitem [{\citenamefont {Chivukula}\ and\ \citenamefont
  {Walker}(1990)}]{Chivukula:1989qb}%
  \BibitemOpen
  \bibfield  {author} {\bibinfo {author} {\bibfnamefont {R.~S.}\ \bibnamefont
  {Chivukula}}\ and\ \bibinfo {author} {\bibfnamefont {T.~P.}\ \bibnamefont
  {Walker}},\ }\href {\doibase 10.1016/0550-3213(90)90151-3} {\bibfield
  {journal} {\bibinfo  {journal} {Nucl.Phys.}\ }\textbf {\bibinfo {volume}
  {B329}},\ \bibinfo {pages} {445} (\bibinfo {year} {1990})}\BibitemShut
  {NoStop}%
\bibitem [{\citenamefont {Barr}\ \emph {et~al.}(1990)\citenamefont {Barr},
  \citenamefont {Chivukula},\ and\ \citenamefont {Farhi}}]{Barr:1990ca}%
  \BibitemOpen
  \bibfield  {author} {\bibinfo {author} {\bibfnamefont {S.~M.}\ \bibnamefont
  {Barr}}, \bibinfo {author} {\bibfnamefont {R.~S.}\ \bibnamefont {Chivukula}},
  \ and\ \bibinfo {author} {\bibfnamefont {E.}~\bibnamefont {Farhi}},\ }\href
  {\doibase 10.1016/0370-2693(90)91661-T} {\bibfield  {journal} {\bibinfo
  {journal} {Phys.Lett.}\ }\textbf {\bibinfo {volume} {B241}},\ \bibinfo
  {pages} {387} (\bibinfo {year} {1990})}\BibitemShut {NoStop}%
\bibitem [{\citenamefont {Barr}(1991)}]{Barr:1991qn}%
  \BibitemOpen
  \bibfield  {author} {\bibinfo {author} {\bibfnamefont {S.~M.}\ \bibnamefont
  {Barr}},\ }\href {\doibase 10.1103/PhysRevD.44.3062} {\bibfield  {journal}
  {\bibinfo  {journal} {Phys.Rev.}\ }\textbf {\bibinfo {volume} {D44}},\
  \bibinfo {pages} {3062} (\bibinfo {year} {1991})}\BibitemShut {NoStop}%
\bibitem [{\citenamefont {Kaplan}(1992)}]{Kaplan:1991ah}%
  \BibitemOpen
  \bibfield  {author} {\bibinfo {author} {\bibfnamefont {D.~B.}\ \bibnamefont
  {Kaplan}},\ }\href {\doibase 10.1103/PhysRevLett.68.741} {\bibfield
  {journal} {\bibinfo  {journal} {Phys.Rev.Lett.}\ }\textbf {\bibinfo {volume}
  {68}},\ \bibinfo {pages} {741} (\bibinfo {year} {1992})}\BibitemShut
  {NoStop}%
\bibitem [{\citenamefont {Lewis}\ \emph {et~al.}(2012)\citenamefont {Lewis},
  \citenamefont {Pica},\ and\ \citenamefont {Sannino}}]{Lewis:2011zb}%
  \BibitemOpen
  \bibfield  {author} {\bibinfo {author} {\bibfnamefont {R.}~\bibnamefont
  {Lewis}}, \bibinfo {author} {\bibfnamefont {C.}~\bibnamefont {Pica}}, \ and\
  \bibinfo {author} {\bibfnamefont {F.}~\bibnamefont {Sannino}},\ }\href
  {\doibase 10.1103/PhysRevD.85.014504} {\bibfield  {journal} {\bibinfo
  {journal} {Phys.Rev.}\ }\textbf {\bibinfo {volume} {D85}},\ \bibinfo {pages}
  {014504} (\bibinfo {year} {2012})},\ \Eprint {http://arxiv.org/abs/1109.3513}
  {arXiv:1109.3513 [hep-ph]} \BibitemShut {NoStop}%
\bibitem [{\citenamefont {Hietanen}\ \emph {et~al.}(2013)\citenamefont
  {Hietanen}, \citenamefont {Pica}, \citenamefont {Sannino},\ and\
  \citenamefont {Sondergaard}}]{Hietanen:2012sz}%
  \BibitemOpen
  \bibfield  {author} {\bibinfo {author} {\bibfnamefont {A.}~\bibnamefont
  {Hietanen}}, \bibinfo {author} {\bibfnamefont {C.}~\bibnamefont {Pica}},
  \bibinfo {author} {\bibfnamefont {F.}~\bibnamefont {Sannino}}, \ and\
  \bibinfo {author} {\bibfnamefont {U.~I.}\ \bibnamefont {Sondergaard}},\
  }\href {\doibase 10.1103/PhysRevD.87.034508} {\bibfield  {journal} {\bibinfo
  {journal} {Phys.Rev.}\ }\textbf {\bibinfo {volume} {D87}},\ \bibinfo {pages}
  {034508} (\bibinfo {year} {2013})},\ \Eprint {http://arxiv.org/abs/1211.5021}
  {arXiv:1211.5021 [hep-lat]} \BibitemShut {NoStop}%
\bibitem [{\citenamefont {Appelquist}\ \emph {et~al.}(2013)\citenamefont
  {Appelquist}, \citenamefont {Brower}, \citenamefont {Buchoff}, \citenamefont
  {Cheng}, \citenamefont {Cohen}, \citenamefont {Fleming}, \citenamefont
  {Kiskis}, \citenamefont {Lin}, \citenamefont {Neil}, \citenamefont {Osborn},
  \citenamefont {Rebbi}, \citenamefont {Schaich}, \citenamefont {Schroeder},
  \citenamefont {Syritsyn}, \citenamefont {Voronov}, \citenamefont {Vranas},\
  and\ \citenamefont {Wasem}}]{Appelquist:2013ms}%
  \BibitemOpen
  \bibfield  {author} {\bibinfo {author} {\bibfnamefont {T.}~\bibnamefont
  {Appelquist}}, \bibinfo {author} {\bibfnamefont {R.~C.}\ \bibnamefont
  {Brower}}, \bibinfo {author} {\bibfnamefont {M.~I.}\ \bibnamefont {Buchoff}},
  \bibinfo {author} {\bibfnamefont {M.}~\bibnamefont {Cheng}}, \bibinfo
  {author} {\bibfnamefont {S.~D.}\ \bibnamefont {Cohen}}, \bibinfo {author}
  {\bibfnamefont {G.~T.}\ \bibnamefont {Fleming}}, \bibinfo {author}
  {\bibfnamefont {J.}~\bibnamefont {Kiskis}}, \bibinfo {author} {\bibfnamefont
  {M.~F.}\ \bibnamefont {Lin}}, \bibinfo {author} {\bibfnamefont {E.~T.}\
  \bibnamefont {Neil}}, \bibinfo {author} {\bibfnamefont {J.~C.}\ \bibnamefont
  {Osborn}}, \bibinfo {author} {\bibfnamefont {C.}~\bibnamefont {Rebbi}},
  \bibinfo {author} {\bibfnamefont {D.}~\bibnamefont {Schaich}}, \bibinfo
  {author} {\bibfnamefont {C.}~\bibnamefont {Schroeder}}, \bibinfo {author}
  {\bibfnamefont {S.~N.}\ \bibnamefont {Syritsyn}}, \bibinfo {author}
  {\bibfnamefont {G.}~\bibnamefont {Voronov}}, \bibinfo {author} {\bibfnamefont
  {P.}~\bibnamefont {Vranas}}, \ and\ \bibinfo {author} {\bibfnamefont
  {J.}~\bibnamefont {Wasem}} (\bibinfo {collaboration} {Lattice Strong Dynamics
  (LSD) Collaboration}),\ }\href {\doibase 10.1103/PhysRevD.88.014502}
  {\bibfield  {journal} {\bibinfo  {journal} {Phys.Rev.}\ }\textbf {\bibinfo
  {volume} {D88}},\ \bibinfo {pages} {014502} (\bibinfo {year} {2013})},\
  \Eprint {http://arxiv.org/abs/1301.1693} {arXiv:1301.1693 [hep-ph]}
  \BibitemShut {NoStop}%
\bibitem [{\citenamefont {Hietanen}\ \emph {et~al.}(2014)\citenamefont
  {Hietanen}, \citenamefont {Lewis}, \citenamefont {Pica},\ and\ \citenamefont
  {Sannino}}]{Hietanen:2013fya}%
  \BibitemOpen
  \bibfield  {author} {\bibinfo {author} {\bibfnamefont {A.}~\bibnamefont
  {Hietanen}}, \bibinfo {author} {\bibfnamefont {R.}~\bibnamefont {Lewis}},
  \bibinfo {author} {\bibfnamefont {C.}~\bibnamefont {Pica}}, \ and\ \bibinfo
  {author} {\bibfnamefont {F.}~\bibnamefont {Sannino}},\ }\href {\doibase
  10.1007/JHEP12(2014)130} {\bibfield  {journal} {\bibinfo  {journal} {JHEP}\
  }\textbf {\bibinfo {volume} {1412}},\ \bibinfo {pages} {130} (\bibinfo {year}
  {2014})},\ \Eprint {http://arxiv.org/abs/1308.4130} {arXiv:1308.4130
  [hep-ph]} \BibitemShut {NoStop}%
\bibitem [{\citenamefont {Appelquist}\ \emph {et~al.}(2014)\citenamefont
  {Appelquist}, \citenamefont {Berkowitz}, \citenamefont {Brower},
  \citenamefont {Buchoff}, \citenamefont {Fleming}, \citenamefont {Kiskis},
  \citenamefont {Kribs}, \citenamefont {Lin}, \citenamefont {Neil},
  \citenamefont {Osborn}, \citenamefont {Rebbi}, \citenamefont {Rinaldi},
  \citenamefont {Schaich}, \citenamefont {Schroeder}, \citenamefont {Syritsyn},
  \citenamefont {Voronov}, \citenamefont {Vranas}, \citenamefont {Weinberg},\
  and\ \citenamefont {Witzel}}]{Appelquist:2014jch}%
  \BibitemOpen
  \bibfield  {author} {\bibinfo {author} {\bibfnamefont {T.}~\bibnamefont
  {Appelquist}}, \bibinfo {author} {\bibfnamefont {E.}~\bibnamefont
  {Berkowitz}}, \bibinfo {author} {\bibfnamefont {R.~C.}\ \bibnamefont
  {Brower}}, \bibinfo {author} {\bibfnamefont {M.~I.}\ \bibnamefont {Buchoff}},
  \bibinfo {author} {\bibfnamefont {G.~T.}\ \bibnamefont {Fleming}}, \bibinfo
  {author} {\bibfnamefont {J.}~\bibnamefont {Kiskis}}, \bibinfo {author}
  {\bibfnamefont {G.~D.}\ \bibnamefont {Kribs}}, \bibinfo {author}
  {\bibfnamefont {M.}~\bibnamefont {Lin}}, \bibinfo {author} {\bibfnamefont
  {E.~T.}\ \bibnamefont {Neil}}, \bibinfo {author} {\bibfnamefont {J.~C.}\
  \bibnamefont {Osborn}}, \bibinfo {author} {\bibfnamefont {C.}~\bibnamefont
  {Rebbi}}, \bibinfo {author} {\bibfnamefont {E.}~\bibnamefont {Rinaldi}},
  \bibinfo {author} {\bibfnamefont {D.}~\bibnamefont {Schaich}}, \bibinfo
  {author} {\bibfnamefont {C.}~\bibnamefont {Schroeder}}, \bibinfo {author}
  {\bibfnamefont {S.}~\bibnamefont {Syritsyn}}, \bibinfo {author}
  {\bibfnamefont {G.}~\bibnamefont {Voronov}}, \bibinfo {author} {\bibfnamefont
  {P.}~\bibnamefont {Vranas}}, \bibinfo {author} {\bibfnamefont
  {E.}~\bibnamefont {Weinberg}}, \ and\ \bibinfo {author} {\bibfnamefont
  {O.}~\bibnamefont {Witzel}} (\bibinfo {collaboration} {Lattice Strong
  Dynamics (LSD) Collaboration}),\ }\href {\doibase 10.1103/PhysRevD.89.094508}
  {\bibfield  {journal} {\bibinfo  {journal} {Phys.Rev.}\ }\textbf {\bibinfo
  {volume} {D89}},\ \bibinfo {pages} {094508} (\bibinfo {year} {2014})},\
  \Eprint {http://arxiv.org/abs/1402.6656} {arXiv:1402.6656 [hep-lat]}
  \BibitemShut {NoStop}%
\bibitem [{\citenamefont {Detmold}\ \emph
  {et~al.}(2014{\natexlab{a}})\citenamefont {Detmold}, \citenamefont
  {McCullough},\ and\ \citenamefont {Pochinsky}}]{Detmold:2014qqa}%
  \BibitemOpen
  \bibfield  {author} {\bibinfo {author} {\bibfnamefont {W.}~\bibnamefont
  {Detmold}}, \bibinfo {author} {\bibfnamefont {M.}~\bibnamefont {McCullough}},
  \ and\ \bibinfo {author} {\bibfnamefont {A.}~\bibnamefont {Pochinsky}},\
  }\href {\doibase 10.1103/PhysRevD.90.115013} {\bibfield  {journal} {\bibinfo
  {journal} {Phys.Rev.}\ }\textbf {\bibinfo {volume} {D90}},\ \bibinfo {pages}
  {115013} (\bibinfo {year} {2014}{\natexlab{a}})},\ \Eprint
  {http://arxiv.org/abs/1406.2276} {arXiv:1406.2276 [hep-ph]} \BibitemShut
  {NoStop}%
\bibitem [{\citenamefont {Detmold}\ \emph
  {et~al.}(2014{\natexlab{b}})\citenamefont {Detmold}, \citenamefont
  {McCullough},\ and\ \citenamefont {Pochinsky}}]{Detmold:2014kba}%
  \BibitemOpen
  \bibfield  {author} {\bibinfo {author} {\bibfnamefont {W.}~\bibnamefont
  {Detmold}}, \bibinfo {author} {\bibfnamefont {M.}~\bibnamefont {McCullough}},
  \ and\ \bibinfo {author} {\bibfnamefont {A.}~\bibnamefont {Pochinsky}},\
  }\href {\doibase 10.1103/PhysRevD.90.114506} {\bibfield  {journal} {\bibinfo
  {journal} {Phys.Rev.}\ }\textbf {\bibinfo {volume} {D90}},\ \bibinfo {pages}
  {114506} (\bibinfo {year} {2014}{\natexlab{b}})},\ \Eprint
  {http://arxiv.org/abs/1406.4116} {arXiv:1406.4116 [hep-lat]} \BibitemShut
  {NoStop}%
\bibitem [{\citenamefont {Fodor}\ \emph {et~al.}(2015)\citenamefont {Fodor},
  \citenamefont {Holland}, \citenamefont {Kuti}, \citenamefont {Mondal},
  \citenamefont {Nogradi} \emph {et~al.}}]{Fodor:2015eea}%
  \BibitemOpen
  \bibfield  {author} {\bibinfo {author} {\bibfnamefont {Z.}~\bibnamefont
  {Fodor}}, \bibinfo {author} {\bibfnamefont {K.}~\bibnamefont {Holland}},
  \bibinfo {author} {\bibfnamefont {J.}~\bibnamefont {Kuti}}, \bibinfo {author}
  {\bibfnamefont {S.}~\bibnamefont {Mondal}}, \bibinfo {author} {\bibfnamefont
  {D.}~\bibnamefont {Nogradi}},  \emph {et~al.},\ }\href@noop {} {\  (\bibinfo
  {year} {2015})},\ \Eprint {http://arxiv.org/abs/1501.06607} {arXiv:1501.06607
  [hep-lat]} \BibitemShut {NoStop}%
\bibitem [{\citenamefont {Chivukula}\ \emph {et~al.}(1993)\citenamefont
  {Chivukula}, \citenamefont {Cohen}, \citenamefont {Luke},\ and\ \citenamefont
  {Savage}}]{Chivukula:1992pn}%
  \BibitemOpen
  \bibfield  {author} {\bibinfo {author} {\bibfnamefont {R.~S.}\ \bibnamefont
  {Chivukula}}, \bibinfo {author} {\bibfnamefont {A.~G.}\ \bibnamefont
  {Cohen}}, \bibinfo {author} {\bibfnamefont {M.~E.}\ \bibnamefont {Luke}}, \
  and\ \bibinfo {author} {\bibfnamefont {M.~J.}\ \bibnamefont {Savage}},\
  }\href {\doibase 10.1016/0370-2693(93)91836-C} {\bibfield  {journal}
  {\bibinfo  {journal} {Phys.Lett.}\ }\textbf {\bibinfo {volume} {B298}},\
  \bibinfo {pages} {380} (\bibinfo {year} {1993})},\ \Eprint
  {http://arxiv.org/abs/hep-ph/9210274} {arXiv:hep-ph/9210274 [hep-ph]}
  \BibitemShut {NoStop}%
\bibitem [{\citenamefont {Bagnasco}\ \emph {et~al.}(1994)\citenamefont
  {Bagnasco}, \citenamefont {Dine},\ and\ \citenamefont
  {Thomas}}]{Bagnasco:1993st}%
  \BibitemOpen
  \bibfield  {author} {\bibinfo {author} {\bibfnamefont {J.}~\bibnamefont
  {Bagnasco}}, \bibinfo {author} {\bibfnamefont {M.}~\bibnamefont {Dine}}, \
  and\ \bibinfo {author} {\bibfnamefont {S.~D.}\ \bibnamefont {Thomas}},\
  }\href {\doibase 10.1016/0370-2693(94)90830-3} {\bibfield  {journal}
  {\bibinfo  {journal} {Phys.Lett.}\ }\textbf {\bibinfo {volume} {B320}},\
  \bibinfo {pages} {99} (\bibinfo {year} {1994})},\ \Eprint
  {http://arxiv.org/abs/hep-ph/9310290} {arXiv:hep-ph/9310290 [hep-ph]}
  \BibitemShut {NoStop}%
\bibitem [{\citenamefont {Pospelov}\ and\ \citenamefont {ter
  Veldhuis}(2000)}]{Pospelov:2000bq}%
  \BibitemOpen
  \bibfield  {author} {\bibinfo {author} {\bibfnamefont {M.}~\bibnamefont
  {Pospelov}}\ and\ \bibinfo {author} {\bibfnamefont {T.}~\bibnamefont {ter
  Veldhuis}},\ }\href {\doibase 10.1016/S0370-2693(00)00358-0} {\bibfield
  {journal} {\bibinfo  {journal} {Phys.Lett.}\ }\textbf {\bibinfo {volume}
  {B480}},\ \bibinfo {pages} {181} (\bibinfo {year} {2000})},\ \Eprint
  {http://arxiv.org/abs/hep-ph/0003010} {arXiv:hep-ph/0003010 [hep-ph]}
  \BibitemShut {NoStop}%
\bibitem [{\citenamefont {Kribs}\ \emph {et~al.}(2010)\citenamefont {Kribs},
  \citenamefont {Roy}, \citenamefont {Terning},\ and\ \citenamefont
  {Zurek}}]{Kribs:2009fy}%
  \BibitemOpen
  \bibfield  {author} {\bibinfo {author} {\bibfnamefont {G.~D.}\ \bibnamefont
  {Kribs}}, \bibinfo {author} {\bibfnamefont {T.~S.}\ \bibnamefont {Roy}},
  \bibinfo {author} {\bibfnamefont {J.}~\bibnamefont {Terning}}, \ and\
  \bibinfo {author} {\bibfnamefont {K.~M.}\ \bibnamefont {Zurek}},\ }\href
  {\doibase 10.1103/PhysRevD.81.095001} {\bibfield  {journal} {\bibinfo
  {journal} {Phys. Rev.}\ }\textbf {\bibinfo {volume} {D81}},\ \bibinfo {pages}
  {095001} (\bibinfo {year} {2010})},\ \Eprint {http://arxiv.org/abs/0909.2034}
  {arXiv:0909.2034 [hep-ph]} \BibitemShut {NoStop}%
\bibitem [{\citenamefont {Buckley}\ and\ \citenamefont
  {Neil}(2013)}]{Buckley:2012ky}%
  \BibitemOpen
  \bibfield  {author} {\bibinfo {author} {\bibfnamefont {M.~R.}\ \bibnamefont
  {Buckley}}\ and\ \bibinfo {author} {\bibfnamefont {E.~T.}\ \bibnamefont
  {Neil}},\ }\href {\doibase 10.1103/PhysRevD.87.043510} {\bibfield  {journal}
  {\bibinfo  {journal} {Phys.Rev.}\ }\textbf {\bibinfo {volume} {D87}},\
  \bibinfo {pages} {043510} (\bibinfo {year} {2013})},\ \Eprint
  {http://arxiv.org/abs/1209.6054} {arXiv:1209.6054 [hep-ph]} \BibitemShut
  {NoStop}%
\bibitem [{\citenamefont {Appelquist}\ \emph {et~al.}(2015)\citenamefont
  {Appelquist}, \citenamefont {Brower}, \citenamefont {Buchoff}, \citenamefont
  {Fleming}, \citenamefont {Jin}, \citenamefont {Kiskis}, \citenamefont
  {Kribs}, \citenamefont {Neil}, \citenamefont {Osborn}, \citenamefont {Rebbi},
  \citenamefont {Rinaldi}, \citenamefont {Schaich}, \citenamefont {Schroeder},
  \citenamefont {Syritsyn}, \citenamefont {Vranas}, \citenamefont {Weinberg},\
  and\ \citenamefont {Witzel}}]{Appelquist:2015yfa}%
  \BibitemOpen
  \bibfield  {author} {\bibinfo {author} {\bibfnamefont {T.}~\bibnamefont
  {Appelquist}}, \bibinfo {author} {\bibfnamefont {R.~C.}\ \bibnamefont
  {Brower}}, \bibinfo {author} {\bibfnamefont {M.~I.}\ \bibnamefont {Buchoff}},
  \bibinfo {author} {\bibfnamefont {G.~T.}\ \bibnamefont {Fleming}}, \bibinfo
  {author} {\bibfnamefont {X.-Y.}\ \bibnamefont {Jin}}, \bibinfo {author}
  {\bibfnamefont {J.}~\bibnamefont {Kiskis}}, \bibinfo {author} {\bibfnamefont
  {G.~D.}\ \bibnamefont {Kribs}}, \bibinfo {author} {\bibfnamefont {E.~T.}\
  \bibnamefont {Neil}}, \bibinfo {author} {\bibfnamefont {J.~C.}\ \bibnamefont
  {Osborn}}, \bibinfo {author} {\bibfnamefont {C.}~\bibnamefont {Rebbi}},
  \bibinfo {author} {\bibfnamefont {E.}~\bibnamefont {Rinaldi}}, \bibinfo
  {author} {\bibfnamefont {D.}~\bibnamefont {Schaich}}, \bibinfo {author}
  {\bibfnamefont {C.}~\bibnamefont {Schroeder}}, \bibinfo {author}
  {\bibfnamefont {S.}~\bibnamefont {Syritsyn}}, \bibinfo {author}
  {\bibfnamefont {P.}~\bibnamefont {Vranas}}, \bibinfo {author} {\bibfnamefont
  {E.}~\bibnamefont {Weinberg}}, \ and\ \bibinfo {author} {\bibfnamefont
  {O.}~\bibnamefont {Witzel}} (\bibinfo {collaboration} {Lattice Strong
  Dynamics (LSD) Collaboration}),\ }\href@noop {} {\  (\bibinfo {year}
  {2015})},\ \Eprint {http://arxiv.org/abs/1503.04203} {arXiv:1503.04203
  [hep-ph]} \BibitemShut {NoStop}%
\bibitem [{\citenamefont {Luke}\ \emph {et~al.}(1992)\citenamefont {Luke},
  \citenamefont {Manohar},\ and\ \citenamefont {Savage}}]{Luke:1992tm}%
  \BibitemOpen
  \bibfield  {author} {\bibinfo {author} {\bibfnamefont {M.~E.}\ \bibnamefont
  {Luke}}, \bibinfo {author} {\bibfnamefont {A.~V.}\ \bibnamefont {Manohar}}, \
  and\ \bibinfo {author} {\bibfnamefont {M.~J.}\ \bibnamefont {Savage}},\
  }\href {\doibase 10.1016/0370-2693(92)91114-O} {\bibfield  {journal}
  {\bibinfo  {journal} {Phys.Lett.}\ }\textbf {\bibinfo {volume} {B288}},\
  \bibinfo {pages} {355} (\bibinfo {year} {1992})},\ \Eprint
  {http://arxiv.org/abs/hep-ph/9204219} {arXiv:hep-ph/9204219 [hep-ph]}
  \BibitemShut {NoStop}%
\bibitem [{\citenamefont {Fitzpatrick}\ \emph {et~al.}(2012)\citenamefont
  {Fitzpatrick}, \citenamefont {Haxton}, \citenamefont {Katz}, \citenamefont
  {Lubbers},\ and\ \citenamefont {Xu}}]{Fitzpatrick:2012ib}%
  \BibitemOpen
  \bibfield  {author} {\bibinfo {author} {\bibfnamefont {A.~L.}\ \bibnamefont
  {Fitzpatrick}}, \bibinfo {author} {\bibfnamefont {W.}~\bibnamefont {Haxton}},
  \bibinfo {author} {\bibfnamefont {E.}~\bibnamefont {Katz}}, \bibinfo {author}
  {\bibfnamefont {N.}~\bibnamefont {Lubbers}}, \ and\ \bibinfo {author}
  {\bibfnamefont {Y.}~\bibnamefont {Xu}},\ }\href@noop {} {\  (\bibinfo {year}
  {2012})},\ \Eprint {http://arxiv.org/abs/1211.2818} {arXiv:1211.2818
  [hep-ph]} \BibitemShut {NoStop}%
\bibitem [{\citenamefont {Fitzpatrick}\ \emph {et~al.}(2013)\citenamefont
  {Fitzpatrick}, \citenamefont {Haxton}, \citenamefont {Katz}, \citenamefont
  {Lubbers},\ and\ \citenamefont {Xu}}]{Fitzpatrick:2012ix}%
  \BibitemOpen
  \bibfield  {author} {\bibinfo {author} {\bibfnamefont {A.~L.}\ \bibnamefont
  {Fitzpatrick}}, \bibinfo {author} {\bibfnamefont {W.}~\bibnamefont {Haxton}},
  \bibinfo {author} {\bibfnamefont {E.}~\bibnamefont {Katz}}, \bibinfo {author}
  {\bibfnamefont {N.}~\bibnamefont {Lubbers}}, \ and\ \bibinfo {author}
  {\bibfnamefont {Y.}~\bibnamefont {Xu}},\ }\href {\doibase
  10.1088/1475-7516/2013/02/004} {\bibfield  {journal} {\bibinfo  {journal}
  {JCAP}\ }\textbf {\bibinfo {volume} {1302}},\ \bibinfo {pages} {004}
  (\bibinfo {year} {2013})},\ \Eprint {http://arxiv.org/abs/1203.3542}
  {arXiv:1203.3542 [hep-ph]} \BibitemShut {NoStop}%
\bibitem [{\citenamefont {Hill}\ and\ \citenamefont
  {Solon}(2014{\natexlab{a}})}]{Hill:2014yka}%
  \BibitemOpen
  \bibfield  {author} {\bibinfo {author} {\bibfnamefont {R.~J.}\ \bibnamefont
  {Hill}}\ and\ \bibinfo {author} {\bibfnamefont {M.~P.}\ \bibnamefont
  {Solon}},\ }\href@noop {} {\  (\bibinfo {year} {2014}{\natexlab{a}})},\
  \Eprint {http://arxiv.org/abs/1401.3339} {arXiv:1401.3339 [hep-ph]}
  \BibitemShut {NoStop}%
\bibitem [{\citenamefont {Hill}\ and\ \citenamefont
  {Solon}(2014{\natexlab{b}})}]{Hill:2014yxa}%
  \BibitemOpen
  \bibfield  {author} {\bibinfo {author} {\bibfnamefont {R.~J.}\ \bibnamefont
  {Hill}}\ and\ \bibinfo {author} {\bibfnamefont {M.~P.}\ \bibnamefont
  {Solon}},\ }\href@noop {} {\  (\bibinfo {year} {2014}{\natexlab{b}})},\
  \Eprint {http://arxiv.org/abs/1409.8290} {arXiv:1409.8290 [hep-ph]}
  \BibitemShut {NoStop}%
\bibitem [{\citenamefont {Frandsen}\ \emph {et~al.}(2012)\citenamefont
  {Frandsen}, \citenamefont {Haisch}, \citenamefont {Kahlhoefer}, \citenamefont
  {Mertsch},\ and\ \citenamefont {Schmidt-Hoberg}}]{Frandsen:2012db}%
  \BibitemOpen
  \bibfield  {author} {\bibinfo {author} {\bibfnamefont {M.~T.}\ \bibnamefont
  {Frandsen}}, \bibinfo {author} {\bibfnamefont {U.}~\bibnamefont {Haisch}},
  \bibinfo {author} {\bibfnamefont {F.}~\bibnamefont {Kahlhoefer}}, \bibinfo
  {author} {\bibfnamefont {P.}~\bibnamefont {Mertsch}}, \ and\ \bibinfo
  {author} {\bibfnamefont {K.}~\bibnamefont {Schmidt-Hoberg}},\ }\href
  {\doibase 10.1088/1475-7516/2012/10/033} {\bibfield  {journal} {\bibinfo
  {journal} {JCAP}\ }\textbf {\bibinfo {volume} {1210}},\ \bibinfo {pages}
  {033} (\bibinfo {year} {2012})},\ \Eprint {http://arxiv.org/abs/1207.3971}
  {arXiv:1207.3971 [hep-ph]} \BibitemShut {NoStop}%
\bibitem [{\citenamefont {Weiner}\ and\ \citenamefont
  {Yavin}(2012)}]{Weiner:2012cb}%
  \BibitemOpen
  \bibfield  {author} {\bibinfo {author} {\bibfnamefont {N.}~\bibnamefont
  {Weiner}}\ and\ \bibinfo {author} {\bibfnamefont {I.}~\bibnamefont {Yavin}},\
  }\href {\doibase 10.1103/PhysRevD.86.075021} {\bibfield  {journal} {\bibinfo
  {journal} {Phys.Rev.}\ }\textbf {\bibinfo {volume} {D86}},\ \bibinfo {pages}
  {075021} (\bibinfo {year} {2012})},\ \Eprint {http://arxiv.org/abs/1206.2910}
  {arXiv:1206.2910 [hep-ph]} \BibitemShut {NoStop}%
\bibitem [{\citenamefont {Ovanesyan}\ and\ \citenamefont
  {Vecchi}(2014)}]{Ovanesyan:2014fha}%
  \BibitemOpen
  \bibfield  {author} {\bibinfo {author} {\bibfnamefont {G.}~\bibnamefont
  {Ovanesyan}}\ and\ \bibinfo {author} {\bibfnamefont {L.}~\bibnamefont
  {Vecchi}},\ }\href@noop {} {\  (\bibinfo {year} {2014})},\ \Eprint
  {http://arxiv.org/abs/1410.0601} {arXiv:1410.0601 [hep-ph]} \BibitemShut
  {NoStop}%
\bibitem [{\citenamefont {Kotila}\ and\ \citenamefont
  {Iachello}(2012)}]{Kotila:2012zza}%
  \BibitemOpen
  \bibfield  {author} {\bibinfo {author} {\bibfnamefont {J.}~\bibnamefont
  {Kotila}}\ and\ \bibinfo {author} {\bibfnamefont {F.}~\bibnamefont
  {Iachello}},\ }\href {\doibase 10.1103/PhysRevC.85.034316} {\bibfield
  {journal} {\bibinfo  {journal} {Phys.Rev.}\ }\textbf {\bibinfo {volume}
  {C85}},\ \bibinfo {pages} {034316} (\bibinfo {year} {2012})},\ \Eprint
  {http://arxiv.org/abs/1209.5722} {arXiv:1209.5722 [nucl-th]} \BibitemShut
  {NoStop}%
\bibitem [{\citenamefont {Mustonen}\ and\ \citenamefont
  {Engel}(2013)}]{Mustonen:2013zu}%
  \BibitemOpen
  \bibfield  {author} {\bibinfo {author} {\bibfnamefont {M.}~\bibnamefont
  {Mustonen}}\ and\ \bibinfo {author} {\bibfnamefont {J.}~\bibnamefont
  {Engel}},\ }\href {\doibase 10.1103/PhysRevC.87.064302} {\bibfield  {journal}
  {\bibinfo  {journal} {Phys.Rev.}\ }\textbf {\bibinfo {volume} {C87}},\
  \bibinfo {pages} {064302} (\bibinfo {year} {2013})},\ \Eprint
  {http://arxiv.org/abs/1301.6997} {arXiv:1301.6997 [nucl-th]} \BibitemShut
  {NoStop}%
\bibitem [{\citenamefont {Detmold}\ \emph {et~al.}(2009)\citenamefont
  {Detmold}, \citenamefont {Tiburzi},\ and\ \citenamefont
  {Walker-Loud}}]{Detmold:2009dx}%
  \BibitemOpen
  \bibfield  {author} {\bibinfo {author} {\bibfnamefont {W.}~\bibnamefont
  {Detmold}}, \bibinfo {author} {\bibfnamefont {B.~C.}\ \bibnamefont
  {Tiburzi}}, \ and\ \bibinfo {author} {\bibfnamefont {A.}~\bibnamefont
  {Walker-Loud}},\ }\href {\doibase 10.1103/PhysRevD.79.094505} {\bibfield
  {journal} {\bibinfo  {journal} {Phys.Rev.}\ }\textbf {\bibinfo {volume}
  {D79}},\ \bibinfo {pages} {094505} (\bibinfo {year} {2009})},\ \Eprint
  {http://arxiv.org/abs/0904.1586} {arXiv:0904.1586 [hep-lat]} \BibitemShut
  {NoStop}%
\bibitem [{\citenamefont {Detmold}\ \emph {et~al.}(2010)\citenamefont
  {Detmold}, \citenamefont {Tiburzi},\ and\ \citenamefont
  {Walker-Loud}}]{Detmold:2010ts}%
  \BibitemOpen
  \bibfield  {author} {\bibinfo {author} {\bibfnamefont {W.}~\bibnamefont
  {Detmold}}, \bibinfo {author} {\bibfnamefont {B.}~\bibnamefont {Tiburzi}}, \
  and\ \bibinfo {author} {\bibfnamefont {A.}~\bibnamefont {Walker-Loud}},\
  }\href {\doibase 10.1103/PhysRevD.81.054502} {\bibfield  {journal} {\bibinfo
  {journal} {Phys.Rev.}\ }\textbf {\bibinfo {volume} {D81}},\ \bibinfo {pages}
  {054502} (\bibinfo {year} {2010})},\ \Eprint {http://arxiv.org/abs/1001.1131}
  {arXiv:1001.1131 [hep-lat]} \BibitemShut {NoStop}%
\bibitem [{\citenamefont {Edwards}\ and\ \citenamefont
  {Joo}(2005)}]{Edwards:2004sx}%
  \BibitemOpen
  \bibfield  {author} {\bibinfo {author} {\bibfnamefont {R.~G.}\ \bibnamefont
  {Edwards}}\ and\ \bibinfo {author} {\bibfnamefont {B.}~\bibnamefont {Joo}}
  (\bibinfo {collaboration} {SciDAC}),\ }\href {\doibase
  10.1016/j.nuclphysbps.2004.11.254} {\bibfield  {journal} {\bibinfo  {journal}
  {Nucl. Phys. Proc. Suppl.}\ }\textbf {\bibinfo {volume} {140}},\ \bibinfo
  {pages} {832} (\bibinfo {year} {2005})},\ \Eprint
  {http://arxiv.org/abs/hep-lat/0409003} {arXiv:hep-lat/0409003} \BibitemShut
  {NoStop}%
\bibitem [{\citenamefont {Bali}\ and\ \citenamefont
  {Bursa}(2008)}]{Bali:2008an}%
  \BibitemOpen
  \bibfield  {author} {\bibinfo {author} {\bibfnamefont {G.~S.}\ \bibnamefont
  {Bali}}\ and\ \bibinfo {author} {\bibfnamefont {F.}~\bibnamefont {Bursa}},\
  }\href {\doibase 10.1088/1126-6708/2008/09/110} {\bibfield  {journal}
  {\bibinfo  {journal} {JHEP}\ }\textbf {\bibinfo {volume} {0809}},\ \bibinfo
  {pages} {110} (\bibinfo {year} {2008})},\ \Eprint
  {http://arxiv.org/abs/0806.2278} {arXiv:0806.2278 [hep-lat]} \BibitemShut
  {NoStop}%
\bibitem [{\citenamefont {Colangelo}\ \emph {et~al.}(2011)\citenamefont
  {Colangelo}, \citenamefont {Durr}, \citenamefont {Juttner}, \citenamefont
  {Lellouch}, \citenamefont {Leutwyler} \emph {et~al.}}]{Colangelo:2010et}%
  \BibitemOpen
  \bibfield  {author} {\bibinfo {author} {\bibfnamefont {G.}~\bibnamefont
  {Colangelo}}, \bibinfo {author} {\bibfnamefont {S.}~\bibnamefont {Durr}},
  \bibinfo {author} {\bibfnamefont {A.}~\bibnamefont {Juttner}}, \bibinfo
  {author} {\bibfnamefont {L.}~\bibnamefont {Lellouch}}, \bibinfo {author}
  {\bibfnamefont {H.}~\bibnamefont {Leutwyler}},  \emph {et~al.},\ }\href
  {\doibase 10.1140/epjc/s10052-011-1695-1} {\bibfield  {journal} {\bibinfo
  {journal} {Eur.Phys.J.}\ }\textbf {\bibinfo {volume} {C71}},\ \bibinfo
  {pages} {1695} (\bibinfo {year} {2011})},\ \Eprint
  {http://arxiv.org/abs/1011.4408} {arXiv:1011.4408 [hep-lat]} \BibitemShut
  {NoStop}%
\bibitem [{\citenamefont {Olive}\ \emph {et~al.}(2014)\citenamefont {Olive}
  \emph {et~al.}}]{Agashe:2014kda}%
  \BibitemOpen
  \bibfield  {author} {\bibinfo {author} {\bibfnamefont {K.}~\bibnamefont
  {Olive}} \emph {et~al.} (\bibinfo {collaboration} {Particle Data Group}),\
  }\href {\doibase 10.1088/1674-1137/38/9/090001} {\bibfield  {journal}
  {\bibinfo  {journal} {Chin.Phys.}\ }\textbf {\bibinfo {volume} {C38}},\
  \bibinfo {pages} {090001} (\bibinfo {year} {2014})}\BibitemShut {NoStop}%
\bibitem [{\citenamefont {Billard}\ \emph {et~al.}(2014)\citenamefont
  {Billard}, \citenamefont {Strigari},\ and\ \citenamefont
  {Figueroa-Feliciano}}]{Billard:2013qya}%
  \BibitemOpen
  \bibfield  {author} {\bibinfo {author} {\bibfnamefont {J.}~\bibnamefont
  {Billard}}, \bibinfo {author} {\bibfnamefont {L.}~\bibnamefont {Strigari}}, \
  and\ \bibinfo {author} {\bibfnamefont {E.}~\bibnamefont
  {Figueroa-Feliciano}},\ }\href {\doibase 10.1103/PhysRevD.89.023524}
  {\bibfield  {journal} {\bibinfo  {journal} {Phys.Rev.}\ }\textbf {\bibinfo
  {volume} {D89}},\ \bibinfo {pages} {023524} (\bibinfo {year} {2014})},\
  \Eprint {http://arxiv.org/abs/1307.5458} {arXiv:1307.5458 [hep-ph]}
  \BibitemShut {NoStop}%
\bibitem [{\citenamefont {Zurek}(2014)}]{Zurek:2013wia}%
  \BibitemOpen
  \bibfield  {author} {\bibinfo {author} {\bibfnamefont {K.~M.}\ \bibnamefont
  {Zurek}},\ }\href {\doibase 10.1016/j.physrep.2013.12.001} {\bibfield
  {journal} {\bibinfo  {journal} {Phys.Rept.}\ }\textbf {\bibinfo {volume}
  {537}},\ \bibinfo {pages} {91} (\bibinfo {year} {2014})},\ \Eprint
  {http://arxiv.org/abs/1308.0338} {arXiv:1308.0338 [hep-ph]} \BibitemShut
  {NoStop}%
\bibitem [{\citenamefont {McDermott}\ \emph {et~al.}(2012)\citenamefont
  {McDermott}, \citenamefont {Yu},\ and\ \citenamefont
  {Zurek}}]{McDermott:2011jp}%
  \BibitemOpen
  \bibfield  {author} {\bibinfo {author} {\bibfnamefont {S.~D.}\ \bibnamefont
  {McDermott}}, \bibinfo {author} {\bibfnamefont {H.-B.}\ \bibnamefont {Yu}}, \
  and\ \bibinfo {author} {\bibfnamefont {K.~M.}\ \bibnamefont {Zurek}},\ }\href
  {\doibase 10.1103/PhysRevD.85.023519} {\bibfield  {journal} {\bibinfo
  {journal} {Phys.Rev.}\ }\textbf {\bibinfo {volume} {D85}},\ \bibinfo {pages}
  {023519} (\bibinfo {year} {2012})},\ \Eprint {http://arxiv.org/abs/1103.5472}
  {arXiv:1103.5472 [hep-ph]} \BibitemShut {NoStop}%
\bibitem [{\citenamefont {Bramante}\ \emph {et~al.}(2013)\citenamefont
  {Bramante}, \citenamefont {Fukushima},\ and\ \citenamefont
  {Kumar}}]{Bramante:2013hn}%
  \BibitemOpen
  \bibfield  {author} {\bibinfo {author} {\bibfnamefont {J.}~\bibnamefont
  {Bramante}}, \bibinfo {author} {\bibfnamefont {K.}~\bibnamefont {Fukushima}},
  \ and\ \bibinfo {author} {\bibfnamefont {J.}~\bibnamefont {Kumar}},\ }\href
  {\doibase 10.1103/PhysRevD.87.055012} {\bibfield  {journal} {\bibinfo
  {journal} {Phys.Rev.}\ }\textbf {\bibinfo {volume} {D87}},\ \bibinfo {pages}
  {055012} (\bibinfo {year} {2013})},\ \Eprint {http://arxiv.org/abs/1301.0036}
  {arXiv:1301.0036 [hep-ph]} \BibitemShut {NoStop}%
\bibitem [{\citenamefont {Bertoni}\ \emph {et~al.}(2013)\citenamefont
  {Bertoni}, \citenamefont {Nelson},\ and\ \citenamefont
  {Reddy}}]{Bertoni:2013bsa}%
  \BibitemOpen
  \bibfield  {author} {\bibinfo {author} {\bibfnamefont {B.}~\bibnamefont
  {Bertoni}}, \bibinfo {author} {\bibfnamefont {A.~E.}\ \bibnamefont {Nelson}},
  \ and\ \bibinfo {author} {\bibfnamefont {S.}~\bibnamefont {Reddy}},\ }\href
  {\doibase 10.1103/PhysRevD.88.123505} {\bibfield  {journal} {\bibinfo
  {journal} {Phys.Rev.}\ }\textbf {\bibinfo {volume} {D88}},\ \bibinfo {pages}
  {123505} (\bibinfo {year} {2013})},\ \Eprint {http://arxiv.org/abs/1309.1721}
  {arXiv:1309.1721 [hep-ph]} \BibitemShut {NoStop}%
\end{thebibliography}%

\end{document}